\documentclass[twocolumn,preprintnumbers,showpacs]{revtex4}
\usepackage{amsmath}
\usepackage{graphicx}
\usepackage{hyperref}

\begin{document}
\title{The Impact of Prior Assumptions on Bayesian Estimates of Inflation
Parameters and the Expected Gravitational Waves Signal from
Inflation} 
\author{Wessel
Valkenburg$^a$\footnote{wessel.valkenburg@lapp.in2p3.fr}}\author{Lawrence
M. Krauss$^b$\footnote{krauss@asu.edu}}\author{Jan
Hamann$^a$\footnote{jan.hamann@lapp.in2p3.fr}} 

\affiliation{$^a$LAPTH\footnote{Laboratoire de Physique
Th\'eorique d'Annecy-le-Vieux, UMR5108}, Universit\'e de Savoie and
CNRS, 9 chemin de Bellevue, BP110, F-74941 Annecy-le-Vieux Cedex,
France}

\affiliation{$^b$School of
  Earth and Space Exploration and Department of Physics, Arizona State University, PO Box 87104, Tempe, Arizona 85287-1404, USA\footnote{current address} and Departments of Physics, Case
  Western Reserve University, Cleveland, Ohio, USA}

\preprint{LAPTH-1247/08, arXiv:0804.3390}

\date{12 September 2008}

\begin{abstract}
  There has been much recent discussion, and some confusion, regarding
  the use of existing observational data to estimate the likelihood
  that next-generation cosmic microwave background (CMB) polarization experiments might detect a
  nonzero tensor signal, possibly associated with inflation.  We
  examine this issue in detail here in two different ways: (1) first
  we explore the effect of choice of different parameter priors on the
  estimation of the tensor-to-scalar ratio $r$ and other parameters
  describing inflation, and (2) we examine the Bayesian complexity in
  order to determine how effectively existing data can constrain
  inflationary parameters. We demonstrate that existing data are not
  strong enough to render full inflationary parameter estimates in a
  parametrization- and prior-independent way and that the predicted
  tensor signal is particularly sensitive to different priors.  For
  parametrizations where the Bayesian complexity is comparable to the
  number of free parameters we find that a flat prior on the scale of
  inflation (which is to be distinguished from a flat prior on the
  tensor-to-scalar ratio) leads us to infer a larger, and in fact
  slightly nonzero tensor contribution at 68\% confidence level.
  However, no detection is claimed.  Our results demonstratethat all that is statistically relevant
  at the current time is the (slightly enhanced) upper bound on $r$,
  and we stress that the data remain consistent with $r = 0$.
\end{abstract}
\maketitle
\section{Introduction}
Shortly after its introduction~
\cite{Starobinsky:1980te,Guth:1980zm,Sato:1980yn,Linde:1981mu,Albrecht:1982wi,Linde:1983gd},
inflation was found to produce a nearly flat Gaussian spectrum of
adiabatic density perturbations that could have been the seeds of
observed structure in the
Universe~\cite{Starobinsky:1979ty,Mukhanov:1981xt,Hawking:1982cz,Starobinsky:1982ee,Guth:1982ec,Bardeen:1983qw,Abbott:1984fp}.
The simplest model of inflation is that of a slowly rolling scalar
field~\cite{Steinhardt:1984jj,Salopek:1990jq,Liddle:1994dx}, which
naturally produces a close to flat primordial spectrum. While the
available observations are remarkably consistent with such a spectrum,
unfortunately one can obtain virtually any scalar spectrum by simply
adjusting the shape of the inflaton potential at early times, and
therefore present results are strongly suggestive, but not yet
unimpeachable evidence that inflation actually occurred.

There are other more generic predictions of inflation that could be
subject to testing, however. For example, a single rolling scalar
field during inflation produces perturbations that are very close to
Gaussian. A detection of significant primordial non-Gaussianity in the
cosmic microwave background (CMB) could rule out simple slow-roll
inflation~\cite{Verde:1999ij}.  A second possibility is the fact that
inflation generally produces a spectrum of tensor perturbations, which
could, among other effects, produce an observable $B$-mode
polarization in the CMB~\cite{Kamionkowski:1996zd,Seljak:1996gy},
albeit plagued by
uncertainties~\cite{Mortonson:2007tb,Martineau:2007dj}. Note that
tensor perturbations are not the only source of $B$-mode
polarization~\cite{Bernardeau:1998mw,Zaldarriaga:1998te,deOliveiraCosta:2002rv,Amarie:2005in},
and noninflationary transitions can also produce a similar background
\cite{krauss,jkm}.  Nevertheless, observation of both the scalar
spectrum and the tensor spectrum could at least test the predictions
of slow-roll (SR) inflation, through the consistency relation
\begin{align}
  n_{\rm T}=-r/8, \qquad \alpha_{\rm T}=n_{\rm T} [n_{\rm T}-n_{\rm
    S}+1], \qquad {\rm etc.},
\end{align}
where $n_{\rm T}$ is the tilt of the tensor spectrum, $r$ is the ratio
of the amplitudes of the tensor and scalar spectra, $\alpha_{\rm T}$
is the running of the tensor spectrum and $n_{\rm S}$ is the tilt of
the scalar spectrum.  A tensor spectrum has not been detected so far,
and many future experiments have been proposed to search for a
gravitational waves signal from
inflation~\cite{:2006uk,Oxley:2005dg,Yoon:2006jc,Bock:2006yf,Maffei:2005,Lawrence:2004,MacTavish:2007kh,Ruhl:2004kv,Kogut:2006nd,Polenta:2007}.

With only observations that constrain the scalar spectrum, one might
hope to gain some information on the inflaton
potential~\cite{Turner:1993su,Copeland:1993jj,Copeland:1993ie}.
However the plethora of different models of inflation make such a task
difficult.  Nevertheless, obtaining any information one can on the
potential using the observed scalar perturbations could give
information about the possibility of observing tensor
perturbations. In Ref.~\cite{Lyth:1996im} the following relation
between the change in value of the scalar field $\phi$ and the
tensor-to-scalar ratio $r$, holding deep inside the slow-roll
approximation, was pointed out,
\begin{align}
\frac{1}{m_{\rm P}}\frac{\Delta \phi}{\Delta N}&\simeq\sqrt{\frac{r}{64\pi}},\label{eq:lyth}
\end{align}
where $N$ is the number of $e$-folds the Universe grows {\em during} the
change $\Delta \phi$ of the scalar field. That is, when focusing on
only a small part of the potential, and not necessarily on the whole
duration of inflation, $\Delta N$ can correspond to a number much
smaller than the total number of $e$-folds of inflation,
$N\sim60-70$. Hence, relation~\eqref{eq:lyth} relates the flatness of
the potential to the relative amplitude of tensor
perturbations. Throughout this work we use $G m_{\rm P}^2=\hbar=c=1$.

As the only current probe of the mechanism of inflation is the
observed spectrum of density perturbations in the Universe,
Refs.~\cite{Peiris:2006ug,Easther:2006tv,Peiris:2006sj,Lesgourgues:2007gp,Lesgourgues:2007aa,Hamann:2008pb}
concentrated on reconstructing the inflaton potential only in the
observational range. It was found in Ref.~\cite{Lesgourgues:2007aa}
that in the observational range naturally $\Delta \phi < m_{\rm P}$
and $\Delta N\sim22$. This bound on $\Delta N$ comes from the
condition that the smallest observable modes actually freeze
in~\cite{Hamann:2008pb}.  The bound on $\Delta \phi$ can then be
understood from Eq.~\eqref{eq:lyth} as the data prefer models with $r$
smaller than at most $0.4$ (depending on the data used).

The reconstruction of the inflation potential gave a weak upper limit
on $r$, fully consistent with $r=0$.  More recently, however,
at least one group has claimed that recent data imply a nonzero
lower limit on $r$ \cite{Destri:2007pv}.  

Obviously it is important to clarify this situation, especially when the
results would have such great significance, and when a dedicated
satellite mission to probe for primordial $B$ modes associated with
a nonzero tensor signal, is being considered.  

In cases such as this, it is useful to take a Bayesian approach and to
consider how effective the data really are at constraining
parameters. Thus, one must consider not merely {\em a posteriori}
probability estimations, but also the effect of prior assumptions 
(see \cite{Parkinson:2006ku} for some discussion of this issue). If
the results depend crucially on the latter, then the parameter
estimates one derives from the data must be taken with a grain of
salt.

The purpose of this paper is to explicitly explore precisely this
question at the current time, in order to help solidify expectations
for future measurements of this important and fundamental quantity
arising from inflation. Specifically we first explore to what extent
the priors one assumes in the analysis affect the expected value of
$r$.

One might argue that with little knowledge of the relevant physics, it
is perhaps pointless to argue strongly on behalf of one set of priors
or another at this point.  It does make sense, however, to examine how
robust the conclusions one draws are, under different prior
assumptions.  (See also work to appear by Vaudrevange and colleagues
\cite{vaudrevange:a, vaudrevange:b}.)  This work focuses on the effect
of taking a flat prior on the Hubble factor during inflation, and its
derivatives with respect to the scalar field value $\phi$.  We will
show that a change of parametrization, but not of physical model, in
this case can lead to significantly different bounds on parameters,
some of which may mildly hint at a larger value of $r$ as well.

In this regard we note that in Ref.~\cite{Destri:2007pv}, a lower
bound on the tensor-to-scalar ratio has been found which one might be
tempted to ascribe to a choice of prior. An important difference
between their result and ours however is that their lower bound on the
tensor-to-scalar ratio is caused by a theoretical prior: the models
they allow can only be consistent with today's observed scalar
amplitude and tilt if the tensor-to-scalar ratio is significant. In
the present work however, the prior on allowed models is as broad as
possible, {\em a priori} not ruling out any combination of inflationary
parameters.

Next, in order to explore the general significance of any derived
lower bound on $r$ based on a choice of priors, we examine the
Bayesian complexity parameter associated with the current data.  This
gives a very useful tool to explore how many free parameters the data
can usefully constrain.  As we demonstrate, for many inflationary
parametrizations, the data are currently simply not powerful enough to
add information beyond the prior, for all the parameters, explaining
the prior-dependence of estimates of $r$ that we have found.  Thus, we
argue that existing data at best provide a rough upper bound on $r$, 
rather than providing a robust 
estimate of its posterior probability distribution. 

In Sec.~\ref{sec:priors} we discuss the relation between different
flat priors, and explain how to translate posterior probability
densities from one prior to another. In Sec.~\ref{sec:result} we
apply a flat prior on the value of the Hubble parameter and its
derivatives during inflation, fit it to the data, and discuss the
results for both prior dependence and Bayesian complexity. We conclude
in Sec.~\ref{sec:conclude}.

\section{Priors and posteriors}\label{sec:priors}

When faced with the problem of estimating parameters from data,
Bayesian inference enjoys a great popularity among cosmologists (see
\cite{Trotta:2008qt} for a recent review). An essential ingredient of
any Bayesian inference is the prior distribution, which encodes our
knowledge about these parameters before any data are taken. With a
suitable basis of parameter space $\{x_i\}$ chosen, it is often
tacitly assumed that the prior is \emph{flat} -- signifying our lack
of information about this parameter in the absence of data. In other
words, the prior probability of an interval $\Delta x_i$ to contain
the true value of the $x_i$ is taken to be constant over the entire
domain of definition of parameter space.

However, while in some problems there is a naturally preferred basis
of parameter space, this need not always be the case, and an
alternative, equally well motivated parametrization $\{y_i\}$ may
exist. It is straightforward to show that generally, a prior in
basis $\{x_i\}$ does not correspond to the same prior in basis
$\{y_i\}$. Labeling a prior A on $\{x_i\}$ by $\pi_x^{(A)}$, the
corresponding prior on $\{y_i\}$ is given by
\begin{align}
 \int {\pi}^{(A)}_{x} d^nx&=1\nonumber\\
 &=  \int {\pi}^{(A)}_{x} \left|\frac{dx_i}{dy_j}\right|d^ny\nonumber\\
&\equiv \frac{1}{V_y} \int {\pi}^{(A)}_{y} d^ny,\\
\pi^{(A)}_y(\vec y)&\propto \pi^{(A)}_x \left(\vec x (\vec
y)\right)\left|\frac{dx_i}{dy_j}\right|
,\label{eq:priortranslation} 
\end{align}
where $V_y = \int d^ny$. Hence a flat, noninformative prior in one
basis does not necessarily equal a noninformative prior in another,
making the choice of basis equivalent to the choice of prior, and by
consequence, extending its influence to the posterior and the inferred
parameter constraints, unless the data become informative enough. This
problem was identified in
\cite{Bucher:2004an,Beltran:2005xd,Beltran:2005gr} in the context of
isocurvature models; here we will argue that inflationary parameters,
including estimates of the tensor-to-scalar ratio, can also be
affected.

\subsection{Importance sampling}
If from earlier analyses one knows that the bounds on parameters in
set $\{x_i\}$ have Gaussian-like shapes, and the sets $\{x_i\}$ and
$\{y_i\}$ are nonlinearly related, one can expect that correlations
between parameters in set $\{y_i\}$ are of nontrivial shape. In that
case a Metropolis-Hastings algorithm, which is what we will use later
on, will have difficulty exploring parameter space properly within an
acceptable amount of time. A solution to this problem is importance
sampling, which is the act of picking points according to one
posterior distribution, but transforming the chance of accepting the
point to another posterior distribution. In this way the algorithm
walks through parameter space according to directions in the
'easier-to-explore' $\{x_i\}$-space, but performing the statistics as
if working in $\{y_i\}$-space. The resulting chains will be
distributed according to the prior chosen in $\{y_i\}$-space. Let $A$
denote statistics with a flat prior on $\{x_i\}$, and let $B$ denote
statistics with a flat prior on $\{y_i\}$. In the Metropolis-Hastings
algorithm, the chance of accepting a proposed step is directly related
to the ratio of its posterior and the posterior of the previous point.
Hence a constant multiplicative factor in the posterior is irrelevant,
and we can neglect the volume term in Eq.~\eqref{eq:priortranslation}.
By consequence, any constant prior corresponds to a flat prior, such
that the conversion to be done is
\begin{align}
  \pi^{(B)}_y(\vec y)&=\left|\frac{dy_i}{dx_j}\right|\pi_y^{(A)}(\vec
  y)=\mbox{constant},\\\mathcal{P}^{(B)}(\vec y(\vec
  x))&=\left|\frac{dy_i}{dx_j}\right|\mathcal{P}^{(A)}(\vec
  x).\label{eq:posttrans}
\end{align} 
There are two distinct places in the analysis in which the correction
for the prior can be applied. One option, which we shall refer to as {
\it post-sampling}, is to take the converged chains of an analysis
performed under prior $(A)$, and multiply the weight of each point in
the chain by the Jacobian as in Eq.~\eqref{eq:posttrans}. 

The advantage is that one can post-process readily available chains to
present a new prior, which takes practically no time. A possible
drawback is that the chains, that converged for an analysis under
prior A, may have too few (or no) points in the regions of parameter
space important under prior B.

The second option is explicit importance sampling of the second
distribution, in which one, during the Monte-Carlo process, transforms
the posterior of a point to reflect the correct prior, by applying
Eq.~\eqref{eq:posttrans} before the decision about acceptance of the
point is taken. The advantage is that the convergence statistics will
now be performed for the correct probability density, hence important
regions will have enough points in the chains. A drawback is that the
analysis has to be performed from scratch, which can be time
consuming.

\subsection{Cosmological parameters}
When constraining the parameters of $\Lambda$CDM cosmologies, it is a
popula choice to take flat priors on $\{\Omega_{\rm c} h^2,
\Omega_{\rm b} h^2, \tau, \theta\}$ (the dark matter density, the
baryon density, the optical depth to reionization, and the ratio of
sound horizon to angular diameter distance at decoupling,
respectively) and for the primordial power spectrum a flat prior on
either $\{\ln A_{\rm S}, n_{\rm S}, \alpha_{\rm S}\}$ (the amplitude,
tilt, and running of the spectrum) or $\{\ln A_{\rm S},
\epsilon_{i}\}$, with $\{\epsilon_{i}\}$ some basis of slow-roll (SR)
parameters. In the SR-basis of Hubble-flow parameters, the dynamics of
inflation are hidden in these parameters by
\begin{align}\label{eq:AS}
A_{\rm S}&=\frac{4 H^4_*}{H'^2_* m_{\rm P}^4},\\
\epsilon&=\frac{m_{\rm P}^2}{4 \pi}\left(\frac{H'_*}{H_*}\right)^2,\\
H_{*}&=\frac{m_{\rm P}^2}{2}\sqrt{\pi A_{\rm S} \epsilon}. 
\end{align}
where $H_{\rm Inf}=H_*$, `$_*$' denotes evaluation at the pivot scale,
and $'$ denotes derivation with respect to the field value of the
inflaton. This means that in all these analyses the posterior
distribution of the derived parameter $H_{\rm Inf}$ is obtained with a
nonflat prior.

\begin{figure}
\includegraphics[width=.5\textwidth]{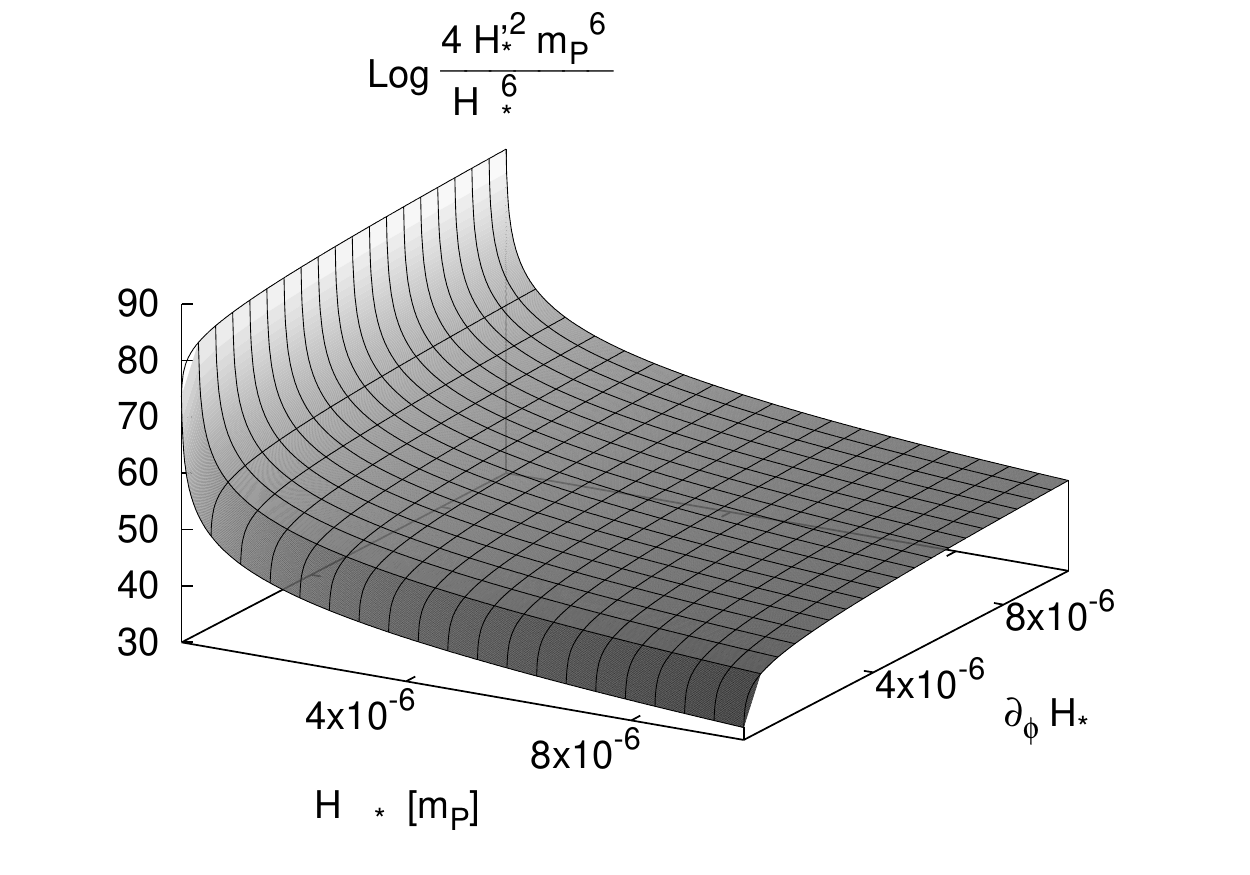}
\caption{The Jacobian $\left|\frac{dx_i}{dy_j}\right|$for the
  coordinate transformation from set $\{x_i\}$ to set $\{y_i\}$.  It
  is clear that a flat prior on set $\{x_i\}$ strongly favors small
  values for $H_*$ compared to a flat prior on set $\{y_i\}$, as
  $\pi_y^{(A)} \sim \pi_x^{(A)} \left|\frac{dx_i}{dy_j}\right|$.
}\label{fig:jac1}
\end{figure}

In Fig.~\ref{fig:jac1} we show the Jacobian
$\left|\frac{dx_i}{dy_j}\right|=\frac{4 H'^2_* m_{\rm P}^6}{H^6_*}$
relating the coordinate transformation between sets
\begin{align}
  \{x_i\}&\equiv\nonumber\\
  &\left\{ \ln \frac{4 H^4_*}{H'^2_* m_{\rm P}^4},
    \left(\frac{H'_*}{H_*}\right)^2 m_{\rm P}^2,
    \frac{H''_*}{H_*}m_{\rm
      P}^2, \frac{H'''_* H'_*}{H_*^2}m_{\rm P}^4 \right\}\\
  \{y_i\}&\equiv\left\{\frac{H_*}{m_{\rm P}}, H'_*, H''_* m_{\rm P},
    H'''_* m_{\rm P}^2\right\},
\end{align} 
corresponding up to a constant to the ratio
$\pi^{(A)}_y/\pi^{(A)}_x$. A flat prior on set $\{y_i\}$ (prior B)
favors high values of $H_*$ when compared to a flat prior on set
$\{x_i\}$ (prior A).

Once the data come into play, the amplitude $A_{\rm S}$ will
essentially be fixed. Since $A_{\rm S} \propto H_*^2/\epsilon$,
higher values of $H_*$ will need to be offset by higher values of
$\epsilon$. In the slow-roll regime, $\epsilon$ is related to the
tensor-to-scalar ratio $r$ by
\begin{align}\label{eq:rsloro}
r&\simeq16\epsilon=\frac{4 m_{\rm P}^2}{\pi}\left(\frac{H'}{H}\right)^2,
\end{align}
and hence we can expect prior B to prefer a larger tensor
contribution, compared to prior A. Equation~\eqref{eq:rsloro} also shows
that a flat prior on $\epsilon$ roughly corresponds to a flat prior on
$r$.

\section{Flat prior on $H_{\rm Inf}$}\label{sec:result}

In order to probe the scale of inflation, we numerically integrate the
perturbation equations of the inflaton in a background described by a
Taylor-expansion of $H(\phi)$, as discussed
in~\cite{Lesgourgues:2007gp,Lesgourgues:2007aa}, and constrain the
free parameters using temperature and polarization data from the five
year data release of the Wilkinson Microwave Anisotropy Probe (WMAP) satellite (WMAP5) \cite{Dunkley:2008ie},
as well as the power spectrum of luminous red galaxies from the Sloan
Digital Sky Survey (SDSS-LRG)~\cite{Tegmark:2006az}. The parameter
estimation is done using the Metropolis-Hastings algorithm, employing
a modified version of the publicly available code
\texttt{CosmoMC}~\cite{Lewis:2002ah} together with our own module for
inflationary perturbations (which is available for download at
\url{http://wwwlapp.in2p3.fr/~valkenbu/inflationH/}).  The parameters
describing the model are either $\{\Omega_{\rm c} h^2, \Omega_{\rm b}
h^2, \tau, \theta\}+\{x_i\}$ or $\{\Omega_{\rm c} h^2, \Omega_{\rm b}
h^2, \tau, \theta\}+\{y_i\}$. We include the calculation of tensor
perturbations.

As a consequence of this exact numerical treatment of perturbations,
we automatically impose a consistent inflationary prior (in the
following this is referred to as ``inflationary consistency''). By
numerically integrating the perturbation equations until the actual
freeze-in of all modes, this method requires inflation to occur over
the observable range, which constrains parameters more strongly than a
naive application of the SR-approximation. As pointed out in
Ref.~\cite{Hamann:2008pb}, a naive implementation of the
SR-approximation allows for inconsistent models, for which small scale
modes actually do not freeze in, even though the approximation
provides a spectrum. We make no prior assumption on the total length
of inflation other than the length needed to produce the observed
power spectrum of perturbations. That is, we remain conservative about
the mechanism of inflation during the unobserved epoch, between
horizon exit of the smallest observable modes and the end of
inflation.
\begin{figure}
\includegraphics[width=.45\textwidth]{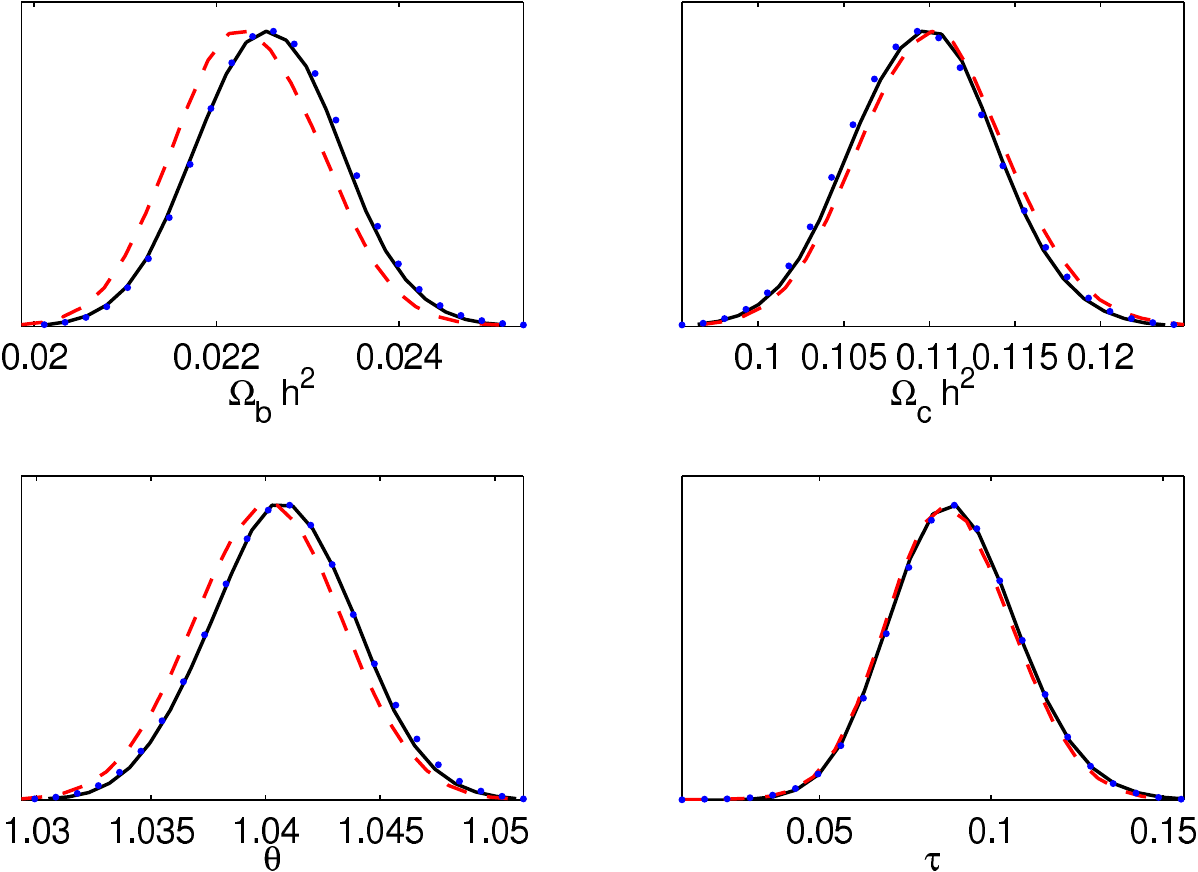}
\caption{The marginalized posterior distributions of cold dark matter
  density ($\Omega_{\rm c} h^2$), the baryon density ($\Omega_{\rm b}
  h^2$), the ratio of sound horizon to angular diameter distance at
  decoupling ($\theta$) and the optical depth to reionization
  ($\tau$), under prior A (red dashed line), post-sampled from prior A
  to prior B (blue dotted line) and under prior B (black solid line).
  The post-sampled distributions are hardly visible as they
  practically completely agree with the importance sampled
  distributions.}\label{fig:fHwm5_cp}
\end{figure}
\begin{figure}
\includegraphics[width=.45\textwidth]{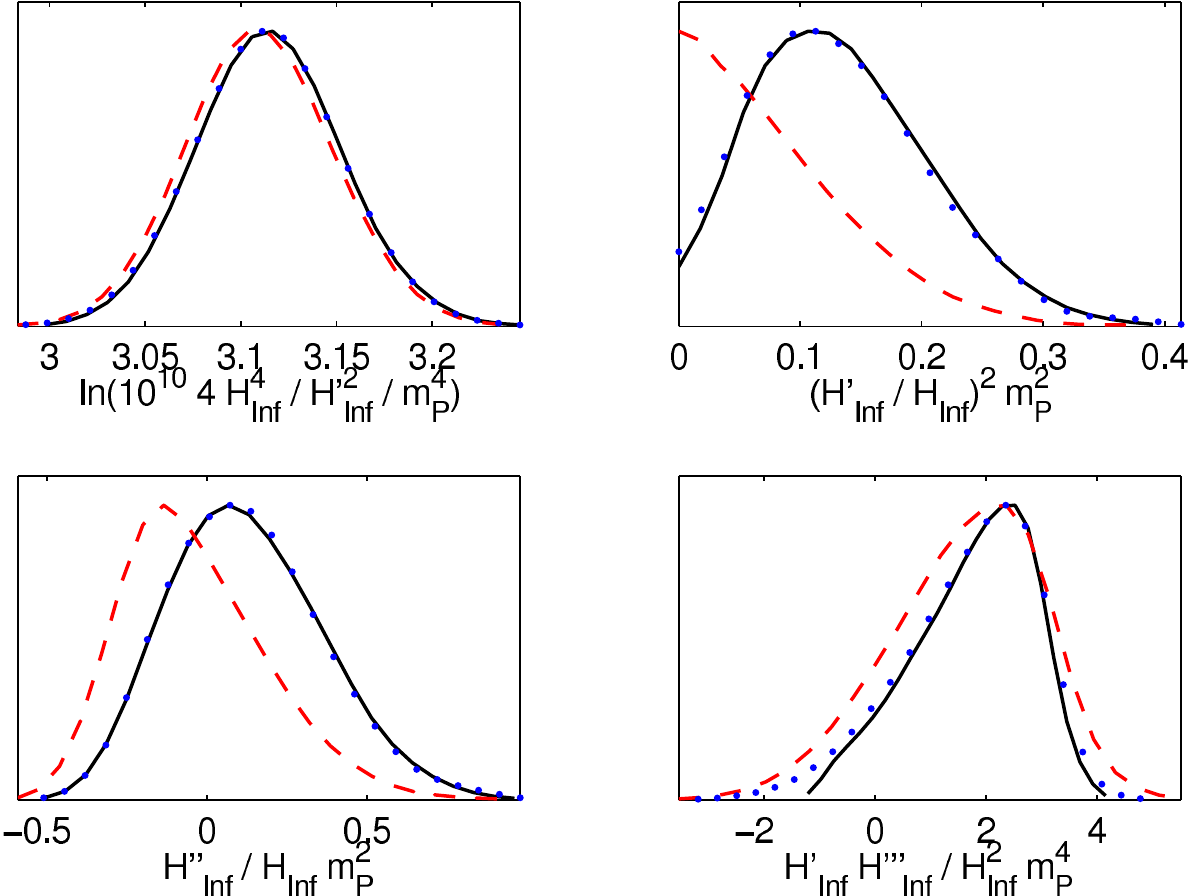}
\caption{The marginalized posterior distributions of the parameters
  describing the evolution of the Universe during inflation, for the
  same analyses as in Fig.~\ref{fig:fHwm5_cp}. Again the post-sampled
  distribution (blue dotted line) is hardly visible due to its good
  agreement with the importance sampled distribution (black solid
  line). The main change under the transformation of priors is seen in
  the posterior of $\left(\frac{H'}{H}\right)^2 m_{\rm P}^2$, in
  agreement with the prediction in
  Fig.~\ref{fig:jac1}.}\label{fig:fHwm5_inf}
\end{figure}
\begin{figure}
\includegraphics[width=.45\textwidth]{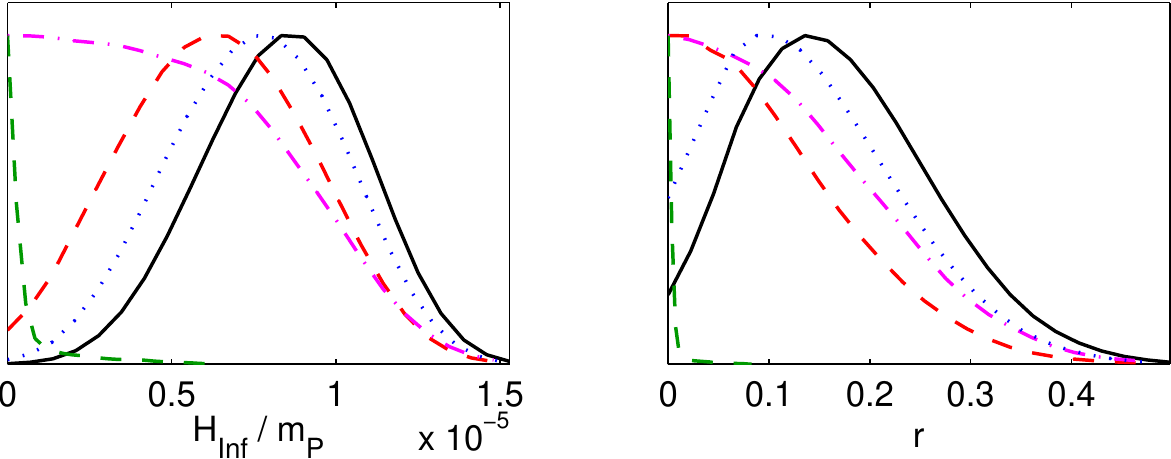}
\caption{The marginalized posterior distributions for the scale of
  inflation, $H_{\rm Inf}$, and the scalar-to-tensor ratio, $r$, under
  prior A (red dashed line), under prior B (black solid lide),
  post-sampled to a Jeffreys prior on $H_*$ (thin blue dotted line)
  and under a flat prior on $\{A_s, \ln\epsilon,  \frac{H''_*}{H_*}m_{\rm
      P}^2, \frac{H'''_* H'_*}{H_*^2}m_{\rm P}^4\}$ (green
  dashed, close to zero for both figures). Prior B corresponds to a
  flat prior on $H_{\rm Inf}$, whereas prior A roughly corresponds to
  a flat prior on $r$, as explained in the text. Prior B pushes both
  $H_{\rm Inf}$ and $r$ up in value. Also shown is the mean likelihood
  over each $(8-1)$--dimensional parameter space for all values of $H$
  and $r$ (dashed-dotted, magenta).}\label{fig:fHwm5_der}
\end{figure}
\begin{figure}
\includegraphics[width=.48\textwidth]{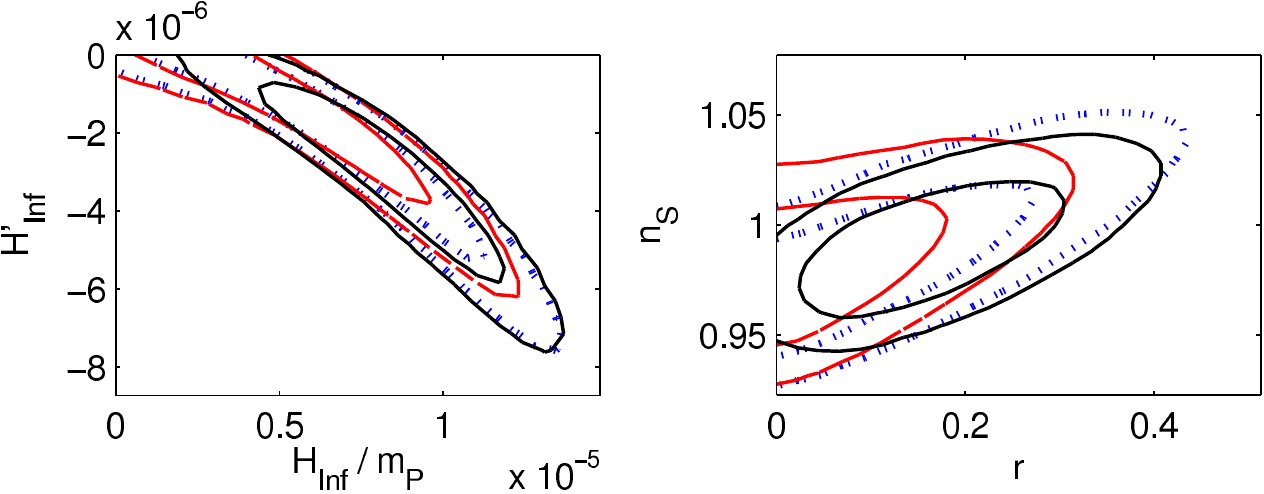}
\caption{Two dimensional marginalized posterior distributions for two
  illustrative cases, comparing prior A (red dashed line), prior B
  (black solid) and a noninflationary analysis, probing the four
  cosmological parameters plus the set $\{\ln A_{\rm S}, n_{\rm S},
  \alpha_{\rm S},r\}$ describing the primordial spectrum (blue dotted
  line). All inner contours correspond to 68\% CL bounds, all outer
  contours correspond to 95\% CL bounds. {\em Left:} the curved
  correlation shape between $\partial_\phi H_{\rm Inf}$ and $H_{\rm
    Inf}$ illustrates the need for importance sampling when taking a
  flat prior on these parameters. {\em Right:} both $r$ and $n_{\rm
    S}$ are pushed up by prior B. }\label{fig:fHwm5_2D}
\end{figure}

As a consistency check we performed both described methods,
post-sampling and importance sampling. In Fig.~\ref{fig:fHwm5_cp} we
show the one dimensional marginalized posterior distributions of the
four cosmological parameters describing the physics after inflation,
comparing the analyses with prior A, the post-sampled chains from
prior A to prior B, and the chains with prior B (importance sampled).
The post-sampled and importance sampled analyses completely agree,
which shows that the chains that converged under prior A have enough
samples in the typical set of the posterior distribution under prior
B. As should be expected, the four parameters shown in
Fig.~\ref{fig:fHwm5_cp} are not affected by the change in prior. In
Fig.~\ref{fig:fHwm5_inf} we show the posterior distributions of
parameters describing the inflationary evolution. The main change is
in the posterior of the parameter $\left(\frac{H'}{H}\right)^2 m_{\rm
  P}^2$, which has a higher preferred value under prior B than under
prior A. This result is in agreement with the expected effect,
illustrated in Fig.~\ref{fig:jac1}.

The scale of inflation and the tensor-to-scalar ratio are shown in
Fig.~\ref{fig:fHwm5_der}. Both parameters, which are related, have a
higher preferred value under prior B.

For illustration, in Fig.~\ref{fig:fHwm5_der}, we also show a
post-sampled distribution with a flat prior on $\left\{\ln H_*, H'_*,
  H''_*, H'''_*\right\}$.  Also this prior gives a higher value for
$H_*$ and $r$ than prior A does. Note that under both prior A and
prior B, we find a lower $r$ than the combined analysis of the WMAP
three year data and SDSS-LRG in Ref.~\cite{Tegmark:2006az}, which is
due to our inflationary prior: integrating the modes until actual
freeze-in, means demanding inflation for about 22 $e$-folds, which
forces the inflaton potential to be relatively smooth. The smoothness
of the potential pushes $\epsilon$ down and thereby also $r$.
Likewise the scalar tilt $n_{\rm S}$ is pushed toward unity, as is
shown in Fig.~\ref{fig:fHwm5_2D} where the two dimensional parameter
correlations are shown for two illustrative cases, $H_{\rm Inf}$
versus $\partial_\phi H$, and $r$ versus $n_{\rm S}$. 
The former illustrates the nonlinear correlation between the
parameters $H_{\rm Inf}$ and $\partial_\phi H$ in the data.  The
curved shape of the posterior probability contour indicates that it
would take a Metropolis-Hastings sampler a long time to random-walk
from one lobe to another if steps are only to be taken in either
horizontal or vertical direction, or a linear combination of both, in
the plane of this plot. By using importance sampling, steps are taken
in correlated directions, significantly speeding up the process.
The latter shows both the effect of the inflationary prior, present in
both analyses, and the effect of going from prior A to prior B. In
both analyses the value of $\epsilon$ is relatively close to zero,
however it is larger under prior B.

It is interesting to note that we also find an apparent lower bound on
the scale of inflation, even for a flat prior on $H_*$. In fact, this
phenomenon is related to our choice of prior on $\epsilon$ (or
$H'_*$, under prior B). Let us illustrate the effect in the example of
prior A.  The dislike of the data for a large tensor contribution
leads to an upper bound on $\epsilon$ due to Eq.~(\ref{eq:rsloro}).
In order to reproduce the observed amplitude of fluctuations, $A_{\rm
  S}\sim H_*^2/\epsilon$ implies also an upper bound on $H_*$.
However, as a consequence of the flat prior on $\epsilon$, extremely
small values of $\epsilon$, while certainly allowed by the data, are
assigned an exponentially suppressed probability, with a preference for
$\epsilon$ of the order of magnitude of its upper bound. Since
$A_{\rm S}\sim H_*^2/\epsilon$, we also have a suppression of small
values of $H_*$, with a peak slightly below the upper bound. If we
instead take the prior to be flat on the logarithm of $\epsilon$
(i.e., a Jeffreys prior on $\epsilon$), we do not see such a
suppression. However, the results for the Jeffreys prior on
$\epsilon$ must be interpreted with care as they are highly
dependent on the lower bound. For numerical reasons we took a lower
bound of $\ln \epsilon > -57$. Had we taken an even smaller lower
bound, the lines would be even closer to zero. A similar result can be
anticipated for a flat prior on $\ln H'_*$ in the $\{y_i\}$
parametrization. That this dependence on the lower bound does not
occur under the Jeffreys prior on $H$ but a flat prior on $H'$ is
explained by the same reasoning as the apparent lower bound on $H$.

In addition to the posteriors, Fig.~\ref{fig:fHwm5_der} shows the mean
likelihood over each $(8-1)$--dimensional parameter space for all
values of $H$ and $r$. This is a prior-independent quantity with no
probabilistic information (i.e., it is not a probability density).  It
serves as an approximation for the profile likelihood. The profile
likelihood is the best fit that can be achieved given a certain
parameter value. The discrepancy between the mean likelihood and the
various posteriors indicates that the various lower bounds in the
posteriors are results of either volume effects in the process of
marginalization, the choice of prior, or a combination of both. The
mean likelihood shows that a good fit can even be achieved for very
small values of $r$ and $H_{\rm Inf}$. In fact the best fit we found
lies at $r=4\times10^{-2}$ and $H_{\rm Inf} =4\times10^{-6} m_{\rm
P}$. This certainly does not coincide with the peaks of the posteriors
found for priors A and B.

In Fig.~\ref{fig:fHwm5_2D} we also compare derived parameters from the
different analyses with the bounds obtained on these parameters when
using no inflationary prior and simply fitting a primordial power
spectrum,
\begin{align}
P(k)&=A_{\rm S} \left(\frac{k}{k_*}\right)^{n_{\rm
S}-1+\frac{1}{2}\alpha_{\rm S}\ln{\frac{k}{k_*}}+\ldots},
\end{align}
and a consistent tensor spectrum, described by $r$ and consistency
relations between the tensor spectral tilt ($n_{\rm T}$) and the
scalar parameters, to the same data. Calculating $\{H_{\rm Inf},
H'_{\rm Inf}\}$ from $\{A_{\rm S},n_{\rm S}, \alpha_{\rm S}, r\}$ is
done using the relations given in Ref.~\cite{Lesgourgues:2007gp}. The
curved shape of the correlation between $H'_{\rm Inf}$ and $H_{\rm
  Inf}$ reflects the need for importance sampling. The 95\% confidence level (CL)
contours under prior A correspond to the 95\% CL contours from the
spectral fit for small values of $H'_{\rm Inf}$, whereas the 95\% CL
contours under prior B correspond to the spectral fit for large values
of $H'_{\rm Inf}$. An important conclusion to draw here is that both
priors A and B allow most of parameter space that is allowed by the
spectral fit, which has no inflationary prior.

In the $n_{\rm S}$-$r$--plane, prior A clearly pushes $r$ down with
respect to merely performing a spectral fit because of the demand that
inflation lasts long enough to produce the full observed spectrum
under a flat prior on virtually the same parameters, whereas prior B
pushes $r$ up, in spite of the same condition on the duration of
inflation.

In the absence of clearly favored theoretical models, the beauty of
various priors is, alas, largely in the eye of the beholder.
Nevertheless we emphasize that both priors A and B do not exceed the
spectral limits but do probe practically the whole range allowed
without the requirement of persistent inflation. More interestingly,
the 68\% CL contour for prior B actually yields a nonzero lower
bound for $r$. Marginalized over all parameters, the posterior of $r$
gives, at 68\% CL, $0.061<r<0.243$, however at 95\% CL $r$ is
still consistent with zero.  While this may hint at a nonzero
amplitude of tensor modes, our analysis underscores the prior model
dependence of this result and thus we do not put much stake in it
here.  Nevertheless, it does suggest that future polarization searches
for tensor modes may have a better chance of detection than otherwise
suggested.

\subsection*{Bayesian Complexity}
\begin{table}
\begin{tabular}{l|l|l|l|l}
  &Prior A&Prior B&$\{A_{\rm S},n_{\rm S}, \alpha_{\rm S},
  r\}$&$\{A_s, \ln\epsilon,  \frac{H''_*}{H_*}m_{\rm
      P}^2, $\\&&&&$\frac{H'''_* H'_*}{H_*^2}m_{\rm P}^4\}$\\ \hline 
  $C_b$&$6.98\pm0.03$&$7.80 \pm 0.03$&$7.76\pm0.06$&$6.25\pm0.8$
\end{tabular}
\caption{Bayesian complexity for different choices of
  prior. This number should be compared to the number of free parameters, which is eight for
  all models considered here.}\label{table:complexity} 
\end{table}
When selecting models and priors, a quantity that can distinguish
between models is the Bayesian evidence, which rewards both the
predictivity and the conciseness of a model, and gives preference to
the model with the best balance between the two characteristics. When
the Bayesian evidence cannot distinguish between two models, a
secondary quantity to make the comparison is the Bayesian
complexity~\cite{Spiegelhalter:2002aa,Kunz:2006mc},
\begin{align}
\mathcal{C}_{\rm b}\equiv\overline{\chi^2}-\chi^2(\hat\theta),
\end{align}
where the effective $\chi^2$ is defined as $-2\ln{\mathcal L}$, with
the likelihood $\mathcal{L}$, and $\hat\theta$ denotes the best fit
point, and the overline denotes the mean over the posterior. The
Bayesian complexity measures the information gain when going from the
prior to the posterior, and can be interpreted as a measure of the
number of parameters the data can constrain in a model, or conversely
the number of parameters a model effectively needs to fit the data.
Along the same lines, if the Bayesian complexity is smaller than the
actual number of free parameters of a model, this could be taken as a
sign that the model contains ``unnecessary'' degrees of freedom, i.e.,
parameters on which we do not gain information from the data.

Note that the Bayesian complexity itself contains no information about
the goodness of fit of a model, or the evidence of one model over
another, but gives an additional measure on the number of parameters
in a model that is justified by the data. Without an evidence
calculation, the complexity can still be useful for telling whether
parameters are mostly bounded by either the data or the prior. In this
work we are interested in the question which priors (with the same
underlying model) constrain parameters beyond the constraining power
of the data, and which priors allow the data to give information on
parameters.

The Bayesian evidence cannot be reliably calculated from the Markov-chain Monte Carlo (MCMC)
chains obtained doing the parameter estimation, as these chains
have a lack of information on the tails of the parameter
distributions. An elaborate analysis would be necessary, e.g. using
nested sampling~\cite{Mukherjee:2005wg}. The Bayesian complexity,
however, can be readily calculated from the chains. 
%
%
In Table~\ref{table:complexity}, we show the Bayesian complexity for
the same model under the different priors.

For prior A we find a complexity of $6.98 \pm 0.03$, for eight free
parameters. This indicates that the data do not give any information
on one of the free parameters. Most likely this is due to $
\frac{H'''_* H'_*}{H_*^2}m_{\rm P}^4$ which is more tightly
constrained by imposing inflationary consistency than by the data, as
explained in Ref.~\cite{Hamann:2008pb}. Compared to prior A, prior B
has more volume in regions constrained by the data, increasing the
amount of information gained and pushing up the complexity to $7.80
\pm 0.03$. Compared to that, in the $\{A_s, \ln\epsilon,
\frac{H''_*}{H_*}m_{\rm P}^2, \frac{H'''_* H'_*}{H_*^2}m_{\rm P}^4\}$
basis the opposite happens, as $\epsilon$ is pushed much closer to
zero, such that the data give no new information on this parameter,
decreasing the complexity.  For the ``phenomenological'' parameter set
$\{A_{\rm S},n_{\rm S}, \alpha_{\rm S}, r\}$ with flat priors, no
inflationary consistency is imposed. Therefore, in this basis,
$\alpha_{\rm S}$ has no theoretical prior constraints and can be
constrained by the data.

An increase in the complexities under prior B and the phenomenological
prior compared to prior A should not be taken to mean that these prior
choices are superior.  Indeed, in all cases the complexity value is
less than 8, which is the number of inflationary parameters under
consideration.  Rather, we take this as an indication that the data is
highly sensitive to the choice of parametrization of inflationary
models, in particular the choice of prior distribution for $r$, and
hence the posterior probability densities reflect far more the choice
of volume of prior parameter space than the impact of the data.

\section{Conclusion}\label{sec:conclude}
Our paper makes explicit an important and well-known fact regarding
the effort to constrain cosmological parameters: the importance of prior
assumptions in the analysis must not be neglected.  We have
demonstrated in a variety of ways that this situation is relevant to
the current issue of a possible nonzero value of $r$ and expectations
for future CMB missions.  In the absence of clear theoretical
direction, it is important therefore to consider the divergence of
results obtained by presumably equally well motivated priors.  We have
demonstrated here how to relate flat priors on different
parametrizations of the same physics, and applied a change of
parametrization to the reconstruction of the inflaton potential,
choosing a flat prior on the parameters that may be better motivated
by the physics of inflation, as opposed to parameters describing the
observable quantities. The main change, seen in
Figs.~\ref{fig:fHwm5_der}~and~\ref{fig:fHwm5_2D}, is an increase of
the preferred value for the tensor-to-scalar ratio $r$, moving from $0
< r < 0.18$ to $0.061<r<0.243$ at $68\%$ CL
We stress once again that this new preferred range does not imply that
the data now prefer a nonzero value of $r$, since at 95\% CL $r$
is consistent with zero under all used priors.  The fact that for
certain choices of parametrization the complexity is less than the
number of free parameters, eight, indicates that the data is currently
insufficient to fully constrain the models.  Rather, our calculation
of the complexity shows that for prior B, which gives the increased
range in $r$, the data are simply sensitive to more of the parameter
volume.  Thus we consider the mean likelihood to be a more meaningful
quantity here.  In particular information on the parameters $r$ and
$H_{\rm Inf}$ under prior B is primarily gained on the upper bound.

As a result we emphasize that the mean likelihood for the parameters
we considered gives an indication of neither a nonzero scale of
inflation nor a nonzero tensor-to-scalar ratio.  Nevertheless, the
fact that one plausible parametrization of the data increases the
posterior probability of these quantities to be nonzero, suggests
that from a Bayesian point of view
the motivation for probing for tensor modes may be slightly enhanced
as a result of our analysis.

\section*{Acknowledgements}
WV acknowledges the hospitality of Case Western Reserve University,
where this work was begun during a very pleasant stay. WV is supported
by the EU 6th Framework Marie Curie Research and Training network
``UniverseNet'' (MRTN-CT-2006-035863). Numerical simulations were
performed on the MUST cluster at LAPP (CNRS \& Universit\'e de
Savoie). LMK is supported by grants from the US DOE and NASA, and
acknowledges the hospitality of the Institute for Theoretical Physics
at University of Zurich, where this work was completed. It is a
pleasure to thank Andrew Liddle, David Parkinson, Hiranya Peiris and
Roberto Trotta for helpful comments and discussions.  We also thank
Pascal Vaudrevange for informing us of his related work in this area.

\bibliography{refs}
\end{document}